\documentclass[aps,prd,
preprintnumbers,nofootinbib,superscriptaddress]{revtex4}
\usepackage[dvips]{graphicx}
\usepackage{caption}
\usepackage{subcaption}
\usepackage{bm,latexsym,amsmath,amssymb,amsfonts,color}
\begin{document}

\title{Universal self-gravitating skyrmions}

\author{Aldo Vera}
\email{aldo.vera@umayor.cl}
\affiliation{N\'ucleo de Matem\'aticas, F\'isica y Estad\'istica, Universidad Mayor, Avenida Manuel Montt 367, Santiago, Chile.}
\affiliation{Centro Multidisciplinario de F\'isica, Vicerrector\'ia de Investigaci\'on, Universidad Mayor, Camino La Pir\'amide 5750, Santiago, Chile.}

\begin{abstract}
The self-gravitating skyrmion is an exact solution of the Einstein $SU(2)$-Skyrme model describing a topological soliton with baryon number $B=1$, living in a $4$-dimensional space-time in the presence of a cosmological constant. Here we show that, using the maximal embedding Ansatz of $SU(2)$ into $SU(N)$ in the Euler angles parametrization, this solution can be generalized to include arbitrary values of the flavor number and, consequently, allowing higher values of the topological charge. Also, we show that higher-order corrections in the 't Hooft expansion can be considered while still preserving the analytical nature of the solutions. Finally we will show that from the gravitational solutions it is possible to construct skyrmions in flat space-time at a finite volume.
\end{abstract}

\maketitle

\newpage 


\section{Introduction}

The Skyrme model is one of the most important non-linear effective field theories that currently exist \cite{Skyrme1}, \cite{Skyrme2}. Constructed from a scalar field $U$ taking values in the $SU(N)$
Lie group (being $N$ the flavor number), it corresponds to the low-energy sector of Quantum Chromodynamics (QCD) at the leading order in the ’t Hooft expansion (when the color number $N_c$ is large) and allows one to describe fermionic degrees of freedom, such as nucleons, both theoretically and phenomenologically \cite{Witten:1979kh}-\cite{Balachandran:1982cb}. In this context, fermions are topological solitons, called skyrmions, and the topological charge is the baryon number of a given field configuration. The static and dynamical properties of skyrmions have been extensively studied in the literature, and the predictions agree well with experimental data \cite{ANW}, \cite{tHooft:1973alw} (see Refs. \cite{MantonBook}- \cite{BalaBook} for details).

When the Skyrme model is coupled to theories of gravity and, in particular, to general relativity in $D=4$ space-time dimensions, it allows modeling gravitating configurations supported by hadronic matter, which are not only interesting from a theoretical point of view but are fundamental in astrophysics. Some relevant pioneering examples are the hairy black hole solutions found in Refs. \cite{Luckock:1986tr} and \cite{Droz:1991cx} (which are also stable under linear perturbations \cite{Heusler:1992av}), and the gravitating solitons constructed in Refs. \cite{Glendenning:1988qy}-\cite{Bizon2}. In recent times novel solutions have been derived, involving black holes with different topologies, wormholes, extended objects, and compact stars (see \cite{Kunz}-\cite{toroidal3} and references therein). However, to solve the field equations of the Einstein-Skyrme system, it is usually necessary to use numerical and approximate resolution methods (see  \cite{Houghton:1997kg}, \cite{Battye:2001qn} and references therein) or accept that the topological charge vanishes (in such a way that the matter content becomes bosonic).

The first topologically non-trivial exact solution of the Einstein-Skyrme model was found in Ref. \cite{Eloy1}. It describes a self-gravitating skyrmion; a regular solution of the Einstein $SU(2)$-Skyrme model with baryon number $B=1$, which lives in a $4$-dimensional space-time in the presence of a cosmological constant (see Ref. \cite{Eloy2} for the extension to the $SU(3)$ symmetry group case). This is the generalization of the skyrmion in a curved space-time reported in Ref. \cite{Manton:1986pz} some decades earlier. In the case of the self-gravitating skyrmion not only the Skyrme equations are solved (as happens in the static hedgehog solution in Ref. \cite{Manton:1986pz}), but also the Einstein equations taking into account the back-reaction of the skyrmion on the space-time geometry.\footnote{The key point in the construction of the solutions in Refs. \cite{Eloy1} and \cite{Eloy2} is to consider a metric compatible with the symmetries of the energy-momentum tensor associated with the $U$ field, even when the $U$ field itself is not invariant under such a symmetry group. Thus, the metric is obtained from the matter field, and not inversely (as is usual in the literature). The same technique has been used to find other regular and cosmological solutions, as can be seen in Refs. \cite{Pavluchenko:2016gyd}-\cite{Canfora:2025roy}.} In addition to the interest that this solution arouses in the area of gravitating solitons -mainly due to its analytical nature-, the techniques developed for its construction have been fundamental for the subsequent development of exact solutions of the Skyrme model in flat space-time at finite volume (see \cite{Pedro1}-\cite{Canfora:2023zmt} and references therein).

At this point, it is important to recall two important issues about the theoretical formulation and experimental results of the Skyrme model. First, the vast majority of solutions of the Skyrme model, both numerical and analytical, have been found in the two-flavors case; that is, when the internal symmetry group is $SU(2)$. Constructing solutions with more flavors than two is generally a complicated task, as can be seen in Refs. \cite{Bala1}-\cite{Ioa} (see also \cite{MantonBook} and \cite{WeinbergBook}). However, in realistic situations, it is necessary to consider the inclusion of more flavors due to some relevant processes being not only restricted to the interaction between pions, neutrons and protons. Second, an optimal description of the low-energy sector of QCD requires considering higher-order terms in the ’t Hooft expansion, so that one should include sub-leading corrections to the Skyrme model.\footnote{In fact, the Skyrme term (which is supplemented to the non-linear sigma model to conform the Skyrme action) is the simplest one that preserves the $SU(N)$ symmetry and the Lorentz invariance, allowing at the same time the existence of topological solitons. However, other terms containing higher derivatives also respect these symmetries and lead to second-order field equations.
} 

In this paper, we will show that the self-gravitating skyrmion constructed in Ref. \cite{Eloy1} can be generalized in two different ways. First, using the maximal embedding Ansatz of $SU(2)$ into $SU(N)$ in the generalized Euler angles parametrization \cite{Bertini:2005rc}-\cite{CacciatoriScotti} (see also Refs. \cite{Tilma:2002ke} and \cite{Tilma:2004kp}), we will promote the solutions to have an arbitrary number of flavors. The main goal here is the derivation of self-gravitating multi-skyrmion states with arbitrarily high baryon number; this without using any approximation. Second, we will show that the integrability of the Einstein-Skyrme equations allowed for the self-gravitating skyrmion is not spoiled when considering the sub-leading corrections to the Skyrme model arising in the ’t Hooft expansion, so that self-gravitating skyrmions are also a solution of the so-called generalized Skyrme model \cite{Marleau:1989fh}-\cite{Marleau:2000ia} (see also Refs. \cite{Jackson:1985yz}-\cite{Gudnason:2017opo}). Finally we will show that from the gravitational solutions it is possible to construct skyrmions in flat space-time at a finite volume.

The paper is organized as follows: In Section \ref{sec-2} we introduce the Einstein-Skyrme model and present a short review of the $SU(2)$ self-gravitating skyrmion. In Section \ref{sec-3} we construct self-gravitating multi-skyrmions for generic $N$ including high-order corrections in the 't Hooft expansion to the model. In addition, we derive skyrmions states at finite volume. Section \ref{sec-4} is dedicated to the conclusions.

\section{Preliminaries} \label{sec-2}

\subsection{The Einstein $SU(N)$-Skyrme model}

The Einstein $SU(N)$-Skyrme model in $D=4$ space-time dimensions with cosmological constant is described by the following action principle
\begin{gather}
I[g, U]=\int_{\mathcal{M}} d^{4}x\sqrt{-g}\left( \frac{R-2\Lambda }{2\kappa }+%
\frac{K}{4}\mathrm{Tr}[L^{\mu }L_{\mu }]+\frac{K\lambda }{32}\mathrm{Tr}%
\left[ G_{\mu \nu }G^{\mu \nu }\right]  \right) \ ,  \label{I} \\ 
L_{\mu }=U^{-1}\nabla_{\mu }U=L_{\mu }^{a} t_{a}\ , \quad G_{\mu \nu }=[L_{\mu},L_{\nu }]\ , \notag
\end{gather}%
where $U(x)\in SU(N)$, $g$ is the metric determinant, $R$ is the Ricci scalar, $\nabla_\mu$ is the Levi-Civita covariant derivative, $\Lambda$ is the cosmological constant, $K$ and $\lambda $ are experimentally fixed positive coupling constants, and $t_{a}$ are the generators of the $SU(N)$ Lie group (or some subgroup of $SU(N)$). Here Greek indices $\{\mu, \nu, \rho...\}$ run over the $4$-dimensional space-time with mostly plus signature, while Latin indices $\{a, b, c, ...\}$ run over the internal space.

The field equations of the system are given by 
\begin{gather}
\nabla_{\mu }L^{\mu }+\frac{\lambda }{4} \nabla_{\mu }[L_{\nu },G^{\mu \nu}]=0\ , \label{Eq1}\\
R_{\mu \nu }-\frac{1}{2}g_{\mu \nu } R+\Lambda g_{\mu \nu }=\kappa T_{\mu \nu} \ , \label{Eq2}  
\end{gather}
where $T_{\mu\nu}$ is the energy-momentum tensor describing the hadronic matter content; it is given by 
\begin{equation}
T_{\mu \nu }=  -\frac{K}{2}\mathrm{Tr}\left[ L_{\mu }L_{\nu }-\frac{1}{2}%
g_{\mu \nu }L^{\alpha }L_{\alpha }\right. +\left. \frac{\lambda }{4}\left(
g^{\alpha \beta }G_{\mu \alpha }G_{\nu \beta }-\frac{1}{4} g_{\mu \nu }
G_{\alpha\beta }G^{\alpha\beta }\right) \right]  \ .
\end{equation}
Note that, in principle, the Skyrme equations are $(N^2-1)$ non-linear coupled second order differential equations.

The topological charge (or winding number), which corresponds to the baryon number in the Skyrme model, is defined by 
\begin{equation}  \label{B}
B=\frac{1}{24\pi^{2}}\int_{\Sigma} \rho_{\text{B}}\, \mathrm{d}V\ , \qquad \rho_{\text{B}}=\epsilon^{ijk}\text{Tr}\biggl[\left( U^{-1}\partial_{i}U\right) \left(
U^{-1}\partial_{j}U\right) \left( U^{-1}\partial_{k}U\right) \biggl]\ \ ,
\end{equation}
where the integration is performed over a space-like hyper-surface $\Sigma$. In the last expression, the indices $\{i, j, k\}$ run in the spatial directions of the space-time.

\subsection{The self-gravitating skyrmion: a short review}

The self-gravitating skyrmion is an exact solution of the Einstein $SU(2)$-Skyrme model. This solution was presented in Ref. \cite{Eloy1}, and it constitutes the generalization of the skyrmion in a curved space-time reported in Ref. \cite{Manton:1986pz}. In the case of the self-gravitating skyrmion not only the equations associated with the matter field are solved (as happens in Ref. \cite{Manton:1986pz}), but also the Einstein equations considering the energy-momentum tensor of the skyrmion as the source. The usual way to write this solution is using the exponential parametrization (or hedgehog Ansatz) for the Skyrme matter field. In such a parametrization, an arbitrary element of $SU(2)$ can be written as 
\begin{gather} \label{U1}
U(x^{\mu })=\cos (\alpha )\mathbf{1}_{2\times 2}+ \sin (\alpha )n^{a}\tau_{a}   \ ,  \\   
n^{1}=\sin \Theta \cos \Phi \ ,\quad n^{2}=\sin \Theta \sin \Phi \ ,\quad n^{3}=\cos \Theta  \ ,  \notag
\end{gather}
where $\alpha=\alpha(x^\mu)$, $\Theta=\Theta(x^\mu)$ and $ \Phi=\Phi(x^\mu)$ are the three degrees of freedom of the $SU(2)$ group. The isospin vectors satisfy $n^a n_a =1 $, with $\tau_a=i \sigma_a$, being $\sigma_a$ the Pauli matrices.

According to Ref. \cite{Eloy1}, when the three degrees of freedom are chosen as
\begin{equation} \label{U2}
 \Phi = \frac{\gamma+\phi}{2} \ , \quad \tan(\Theta)= \frac{\cot(\frac{\theta}{2})}{\cos(\frac{\gamma-\phi}{2})} \ ,
 \quad \tan (\alpha) =\frac{\sqrt{1+\tan^2\Theta}}{\tan(\frac{\gamma-\phi}{2})} \ ,
\end{equation}
and the space-time metric has the form 
\begin{equation} \label{metric1}
    ds^2 = -dt^2 + h(t) [ (d\gamma+\cos(\theta) d \varphi)^2 + d\theta^2 +\sin^2(\theta) d \varphi^2] \ , 
\end{equation}
the Skyrme equations in Eq. \eqref{Eq1} are automatically satisfied, while the Einstein equations in Eq. \eqref{Eq2} are reduced to the following system: 
\begin{gather} \label{Eqh1}
    (\dot{h})^2 - \frac{\Lambda}{3} h^2 - \frac{K \kappa \lambda}{32} \frac{1}{h^2} - \frac{K\kappa -2}{8} = 0 \ , \\ 
    \ddot{h}- \frac{\Lambda}{3} h + \frac{K \kappa \lambda }{32} \frac{1}{h^3} = 0 \ .  \label{Eqh2}
\end{gather}
The above equations are a compatible system for the function $h(t)$ (namely, the scale factor) that can be solved analytically. In fact, Eq. \eqref{Eqh1} is the first integral of Eq. \eqref{Eqh2}, and Eq. \eqref{Eqh2} is a particular case of the Ermakov-Pinney equation.\footnote{Although is clear that Eq. \eqref{Eqh1} implies Eq. \eqref{Eqh2}, we write explicitly both equations to make contact with the static case, which leads to two conditions for the coupling constants (see Eq. \eqref{constraint1}).} See Ref. \cite{Canfora:2017ivv} for the explicit resolution of this system.

In the static case, that is $h(t)=h_0$, the system is solved by a fine-tuning between the coupling constants, that is 
\begin{equation} \label{constraint1}
    h_0^2 = \frac{3(2-K\kappa)}{4\Lambda} \ , \qquad \lambda= \frac{3(2-K\kappa)^2}{8 K \kappa \Lambda} \ . 
\end{equation}
The above configurations (both static and dynamic) describe a topological soliton with unit  baryon charge. In fact, considering Eq. \eqref{B}, and taking into account the following range for the spatial coordinates 
\begin{equation}
    0 \le \gamma < 4\pi \ , \quad 0 \le \theta < \pi \ , \quad 0 \le \varphi < 2\pi \ , 
\end{equation}
the topological charge turns out to be
\begin{equation}
  \rho_{\text{B}}= \frac{3}{2} \sin(\theta) \quad \Rightarrow \quad   B= \frac{1}{24 \pi^2} \int  \rho_{\text{B}} \,  \mathrm{d}\gamma \mathrm{d}\theta \mathrm{d}\varphi = 1 \ . 
\end{equation}
As we have mentioned before, the configuration here reviewed is the first topologically non-trivial solution found in the Einstein-Skyrme model. In the static case, it describes a self-gravitating skyrmion in a space-time with geometry $\mathbb{R}\times S^3$, and where the cosmological constant and the scale factor are fixed in terms of the coupling constants of the model. On the other hand, the dynamical case represents a bouncing state in which the universe contracts and expands, and where a minimum size of the universe exists bounded by the skyrmion scale. In the next section, we will generalize this solution in two different ways. First, we include arbitrary flavors, consequently allowing a higher baryon number. Second, we consider high-order corrections in the 't Hoof expansion to the Skyrme model.

\section{Generalized self-gravitating skyrmions with arbitrary flavors} \label{sec-3}

In order to generalize the above solution including arbitrary flavors and higher order corrections in the 't Hooft expansion, the first step is to realize that the $U$ field that allows the existence of the self-gravitating skyrmion (shown in Eqs. \eqref{U1} and \eqref{U2} in the exponential parametrization) can be easily written using the Euler angles parametrization. Indeed, an arbitrary element of $SU(2)$ can be expressed as
\begin{equation} \label{U3}
 U(x^\mu)= e^{F_1(x^\mu) \cdot \tau_3} e^{F_2(x^\mu) \cdot \tau_2} e^{F_3(x^\mu) \cdot \tau_3}  \ , 
\end{equation}
where $F_1=F_1(x^\mu)$, $F_2=F_2(x^\mu)$ and $F_3=F_3(x^\mu)$ are the three degrees of freedom of $SU(2)$. One can check that the matter field in Eq. \eqref{U2} can be also obtained using the Euler angles parametrization, simply choosing the degrees of freedom as 
\begin{equation} \label{U4}
 F_1(x^\mu) = \gamma \ , \qquad F_2(x^\mu) =  \theta \ , \qquad F_3(x^\mu) = \varphi \ . 
\end{equation}
The above parametrization is very convenient not only from the computation point of view, it also provides a natural way to generalized the self-gravitating skyrmion to include arbitrary number of flavors, as we will see below. 

\subsection{Arbitrary flavors and high baryonic charge}

To construct exact self-gravitating skyrmion configurations with an arbitrary number of flavors, we will use the maximal embedding Ansatz of $SU(2)$ into $SU(N)$ in the generalized Euler angles parametrization. This uses as a base three $N\times N$ matrices, which give rise to an irreducible representation of $SU(2)$ in $SU(N)$ of spin $j=(N-1)/2$. This formalism was introduced in Refs. \cite{Bertini:2005rc}-\cite{CacciatoriScotti} (see also Refs. \cite{Tilma:2002ke} and \cite{Tilma:2004kp}), and this has proven to be very useful in obtaining non-embedded exact solutions, both in Yang-Mills and in the Skyrme model, where the role playing by the $N$ parameter is clearly reveled (see Refs. \cite{CacciatoriScotti} and \cite{Pedro2}-\cite{Gomberoff}).\footnote{Here non-embedded solutions means solutions that cannot be written as trivial embeddings of $SU(2)$ in $SU(N)$.} Such a construction is similar to that used in the pioneering papers \cite{Bala1} and \cite{Bala2}, in the sense that the authors used the $SO(3)$ subgroup of $SU(3)$ to construct di-baryon states in the Skyrme model.

According to Refs. \cite{Bertini:2005rc}-\cite{CacciatoriScotti}, in the maximal embedding Ansatz, the matter field $U(x)\in SU(N)$ in the Euler angles parametrization reads  
\begin{equation}
    U=e^{F_{1}\left(x^{\mu} \right)\cdot T_{3}}e^{F_{2}\left(x^{\mu}\right) \cdot T_{2}} e^{F_{3}\left(x^{\mu}\right) \cdot T_{3}} \ ,  \label{matter-ansatz}
\end{equation}
where the degrees of freedom $F_i$ are arbitrary functions of the space-time coordinates, and  $T_{i}$ (with $i=1,2,3$) are the generators of a $3$-dimensional sub-algebra of $\mathfrak{su}(N)$, which are explicitly given by 
\begin{align}
T_1&=-\frac{i}{2}\sum_{j=2}^{N} \sqrt{(j-1)(N-j+1)}(E_{j-1,j}+E_{j,j-1}) \ ,
\label{T1} \\
T_2&=\frac{1}{2}\sum_{j=2}^{N} \sqrt{(j-1)(N-j+1)}(E_{j-1,j}-E_{j,j-1}) \ ,
\label{T2} \\
T_3&=i\sum_{j=1}^{N} (\frac{N+1}{2}-j)E_{j,j} \ ,  \label{T3}
\end{align}
where $(E_{i,j})_{mn}=\delta_{im}\delta_{jn}$, and being $\delta_{ij}$ the Kronecker delta. These matrices satisfy the following commutation relation
\begin{equation}
\left[T_{a}, T_{b} \right] \  = \  \epsilon_{abc}T_{c} \ . 
 \end{equation}
A clear way to see that configurations constructed using this formalism are non-trivial embeddings of $SU(2)$ into $SU(N)$ is through the computation of the trace of the generators in Eqs. \eqref{T1}, \eqref{T2} and \eqref{T3}, that is
\begin{equation} \label{Tr}
\text{Tr}\left(T_{b}T_{c} \right) \  = \ - \frac{1}{2} a_N\delta_{bc} \ , \qquad a_N = \frac{N\left(N^2-1 \right)}{6} \ . 
\end{equation}
From Eq. \eqref{Tr} we see that the trace of the generators depends explicitly on the $N$ parameter, whose result is proportional to the tetrahedral numbers. This means that matter fields, by considering different values of the flavor number, lead to different solutions with a particular spin (see Refs. \cite{Bertini:2005rc}-\cite{Tilma:2004kp}). We will see below that this fact is fundamental because both the energy density and the topological charge depend on the trace of the generators and, consequently, on the number of flavors.\footnote{For more details about the generalization of the Euler parametrization of $SU(2)$ to any compact Lie group and its application to nuclear physics (and to measure theory in infinite dimensions), we recommend the review in Ref. \cite{CacciatoriScotti}.}

Now, we can come back to our porpoise generalization. Let us again consider the space-time described by the metric in Eq. \eqref{metric1}, but now accompanied by the Skyrme field using the maximal embedding Ansatz of $SU(2)$ into $SU(N)$, that is,
\begin{gather} \label{U5}
 U(x^\mu)= e^{F_1(x^\mu) \cdot T_3} e^{F_2(x^\mu) \cdot T_2} e^{F_3(x^\mu) \cdot T_3}  \ , \\
F_1(x^\mu) = \gamma \ , \qquad F_2(x^\mu) =  \theta \ , \qquad F_3(x^\mu) = \varphi \ ,  \notag
\end{gather}
where the matrices $T_a$ have been defined in Eqs. \eqref{T1}, \eqref{T2} and \eqref{T3}. First, with this Ansatz, the Skyrme equations in Eq. \eqref{Eq1} are still automatically satisfied. The reason behind this is clear; the trace of the generators -which provides ``the $SU(N)$ character'' of the new configurations- is a global factor at the level of the Skyrme equations (that is why the trace of the generators is not written in Eq. \eqref{Eq1}). Second, the Einstein equations in Eq. \eqref{Eq2}, by including the trace of the generators in the energy-momentum tensor, are indeed susceptible to the number of flavors. Thus, the Einstein equations are reduced to the following system for the scale factor:
\begin{gather} \label{Eqh3}
    (\dot{h})^2 - \frac{\Lambda}{3} h^2 - \frac{K \kappa a_N \lambda}{2} \frac{1}{h^2} - \frac{K\kappa a_N}{2} +1 = 0 \ , \\ 
     \ddot{h}- \frac{\Lambda}{3} h + \frac{K \kappa a_N \lambda }{2} \frac{1}{h^3}  = 0 \ , \label{Eqh4}
\end{gather}
where $a_N$ has been defined in Eq. \eqref{Tr}.

In the static case, $h(t)=h_0$, the system is solved by the following fine-tuning between the coupling constants
\begin{equation} \label{constraint2}
    h_0^2 = \frac{3}{2\Lambda} \biggl( 1-\frac{1}{2} a_N K \kappa \biggl) \ , \qquad \lambda = \frac{3}{2 \Lambda a_N K \kappa}\biggl(1-\frac{1}{2}a_N K \kappa  \biggl)^2 \ . 
\end{equation}
The above configurations are the natural generalization of the solutions reported in Refs. \cite{Eloy1} and \cite{Eloy2} (see Section \ref{sec-2}), now including an arbitrary number of flavors labeled by the integer $N$. 
As $\lambda$ is an experimentally fixed positive number, $\Lambda$ must also be positive, and the following constraint for the parameter $N$ must be satisfied: $a_N < 2/(K \kappa)$.

The main relevance of these novel solutions is that, although the field equations are not significantly different for those of the two-flavors case\footnote{Indeed, the system in Eqs. \eqref{Eqh3} and \eqref{Eqh4} can be obtained by making the change $K\rightarrow K a_N$ in Eqs. \eqref{Eqh1} and \eqref{Eqh2}, and then all the results in Ref. \cite{Canfora:2017ivv} can be applied}, the topological charge is not a unit. Instead, it directly depends on the number of flavors considered in the theory. In fact, computing the topological charge from Eq. \eqref{B}, we found
\begin{equation} \label{charge}
    \rho_{\text{B}} = \frac{3}{2} a_N \sin(\theta) \quad \Rightarrow \quad  
    B=\frac{1}{24 \pi^2} \int  \rho_{\text{B}} \,  \mathrm{d}\gamma \mathrm{d}\theta \mathrm{d}\varphi =  a_N \ . 
\end{equation}
We see that the topological charge is proportional to the trace of the generators defined in Eq. \eqref{Tr}. For the first allowed number of flavors; $N=\{2, 3, 4, ...\}$, we obtain $B=\{1, 4, 10, ...\}$; the tetrahedral numbers. The first case corresponds to the self-gravitating skyrmion reported in Ref. \cite{Eloy1}, and the second case corresponds to the $4$-baryon states presented in Ref. \cite{Eloy2}. For large values of $N$, one can describe multi-skyrmion states with an arbitrary high baryon number.

\subsection{Self-gravitating skyrmions in low QCD}

As we have pointed out before, an optimal description of the low-energy sector of QCD requires considering the higher-order terms that come from the 't Hooft expansion when the number of colors is large. In the context of the Skyrme model, this implies supplementing the action with higher-order derivative terms.  This new theory is known as the generalized Skyrme model (see Refs. \cite{Marleau:1989fh}-\cite{Gudnason:2017opo}), which has been studied in the literature both in flat space-time and when the model is coupled to gravity (see Refs. \cite{Floratos}-\cite{Adam:2020yfv} for the construction of some solutions in this model). Of course, obtaining solutions in the generalized model is much more complicated than in the original Skyrme theory. However, as we will see below, the self-gravitating skyrmion is still a solution of the field equations, proving robust to the inclusion of higher-order terms.

The generalized Einstein $SU(N)$-Skyrme model is described by the following action principle
\begin{equation} \label{I2}
I_{\text{gen}}[g, U]=\int_{\mathcal{M}} d^{4}x\sqrt{-g}\left( \frac{R-2\Lambda }{2\kappa }+%
\frac{K}{4}\mathrm{Tr}[L^{\mu }L_{\mu }]+\frac{K\lambda }{32}\mathrm{Tr}%
\left[ G_{\mu \nu }G^{\mu \nu }\right]  \right)+\mathcal{L}_{\text{corr}} \ ,  
\end{equation}%
where $\mathcal{L}_{\text{corr}}$ represents the sub-leading corrections to the Skyrme model. The first two sub-leading terms are given by\footnote{Here we will consider the first two corrections to the Skyrme model, but the construction presented here can be generalized to include other higher order terms (see, for instance, Ref. \cite{SU(N)1}).} 
\begin{align}
\mathcal{L}_{6}=& \frac{c_{6}}{96}\text{Tr}\left[ G_{\mu }{}^{\nu }G_{\nu
}{}^{\rho }G_{\rho }{}^{\mu }\right] \ ,  \label{L6} \\
\mathcal{L}_{8}=& -\frac{c_{8}}{256}\biggl(\text{Tr}\left[ G_{\mu }{}^{\nu
}G_{\nu }{}^{\rho }G_{\rho }{}^{\sigma }G_{\sigma }{}^{\mu }\right] -\text{Tr%
}\left[ \{G_{\mu }{}^{\nu },G_{\rho }{}^{\sigma }\}G_{\nu }{}^{\rho
}G_{\sigma }{}^{\mu }\right] \biggl)\ ,  \label{L8}
\end{align}%
where $c_6$ and $c_8$ are coupling constants. Let us recall that in the Skyrme action the coupling $K$ is fixed in terms of the pions decay constant, $f_\pi$, through the relation $f_\pi^2=4 K$. This ensures that the non-linear sigma model reproduces the properties of pions at low energies. On the other hand, the coupling $\lambda$, which is usually presented as $K \lambda e^2 =1$, can take different values depending on whether one focuses on the properties of a single particle or nuclear matter; see Ref. \cite{BalaBook} for a discussion. Usually is assumed that $e \approx 5.4$ to make contact with the static properties of a nucleon \cite{ANW}, but nowadays its value is still under discussion. Much less clear is the value that the coupling constants of the generalized Skyrme model should take. Here we initially consider that $c_6$ and $c_8$ are arbitrary numbers, although, as we will see below, the existence of self-gravitating skyrmions introduces certain constraints on these constants.

Considering the corrections in Eqs. \eqref{L6} and \eqref{L8}, the field equations of the generalized Skyrme model take the form
\begin{align}
\frac{K}{2}\biggl(\nabla^{\mu }L_{\mu }+\frac{\lambda }{4}\nabla^{\mu }[L^{\nu},G_{\mu
\nu }]\biggl) +3c_{6}[L_{\mu },\nabla_{\nu }[G^{\rho
\nu },G_{\rho }{}^{\mu }]] &  \notag \\
+4c_{8}\biggl[L_{\mu },\nabla_{\nu }\biggl(G^{\nu \rho }G_{\rho \sigma }G^{\sigma
\mu}+G^{\mu \rho }G_{\rho \sigma }G^{\nu \sigma }+\{G_{\rho \sigma
},\{G^{\mu \rho },G^{\nu \sigma }\}\}\biggl)\biggl] & \ = \ 0\ .
\label{Eqgeneralized}
\end{align}%
The Einstein equations in Eq. \eqref{Eq2} must be considered including the contributions to the energy-momentum tensor that comes from the new terms in Eqs. \eqref{L6} and \eqref{L8}. In this case, the full energy-momentum tensor is given by the sum
\begin{equation}  \label{Tmunu2}
T_{\mu \nu }\ =\ T_{\mu \nu }^{\text{Sk}}+T_{\mu
\nu }^{(6)}+T_{\mu \nu }^{(8)} \ ,
\end{equation}%
where the contributions of the Skyrme model, and the higher order corrections associated to $c_6$ and $c_8$ are, respectively,
\begin{align*}
T_{\mu \nu }^{\text{Sk}}=& -\frac{K}{2}\text{Tr}\biggl(L_{\mu }L_{\nu }-%
\frac{1}{2}g_{\mu \nu }L_{\alpha }L^{\alpha }+\frac{\lambda }{4}(g^{\alpha
\beta }G_{\mu \alpha }G_{\nu \beta }-\frac{1}{4}g_{\mu \nu }G_{\alpha \beta
}G^{\alpha \beta })\biggl)\ , \\
T_{\mu \nu }^{(6)}=& -\frac{c_{6}}{16}\text{Tr}\biggl(g^{\alpha \gamma
}g^{\beta \rho }G_{\mu \alpha }G_{\nu \beta }G_{\gamma \rho }-\frac{1}{6}%
g_{\mu \nu }G_{\alpha }{}^{\beta }G_{\beta }{}^{\rho }G_{\rho }{}^{\alpha }%
\biggl)\ , \\
T_{\mu \nu }^{(8)}=& \ \frac{c_{8}}{32}\text{Tr}\biggl(g^{\alpha \rho
}g^{\beta \gamma }g^{\delta \lambda }G_{\alpha \mu }G_{\nu \beta }G_{\gamma
\delta }G_{\lambda \rho }+\frac{1}{2}\{G_{\mu \alpha },G_{\lambda \rho
}\}\{G_{\beta \nu },G_{\gamma \delta }\}g^{\alpha \gamma }g^{\beta \rho
}g^{\delta \lambda }  \notag \\
& \qquad -\frac{1}{8}g_{\mu \nu }(G_{\alpha }{}^{\beta }G_{\beta }{}^{\rho }G_{\rho
}{}^{\sigma }G_{\sigma }{}^{\alpha }-\{G_{\alpha }{}^{\beta },G_{\rho
}{}^{\sigma }\}G_{\beta }{}^{\rho }G_{\sigma }{}^{\alpha })\biggl)\ .
\end{align*}
Now, considering the Ansatz for the matter field and the metric in Eqs. \eqref{metric1} and \eqref{U5}, one can check that the generalized Skyrme equations in Eq. \eqref{Eqgeneralized} are reduced to the following constraint 
\begin{equation*}
    (c_6 h(t)^2 - (N^2-5) c_8) \cot(\theta) = 0   \ , 
\end{equation*}
which implies that the metric is static, and that a fine-tuning between the new coupling constants must be fulfilled: 
\begin{equation} \label{newconst}
    h(t)=h_0 \ , \qquad  c_8= \frac{1}{(N^2-5)} c_6 h_0^2 \ .
\end{equation}
It is also possible to verify that the Einstein equations reduce to two algebraic constraints involving all the constants of the theory, similarly to what happens in the classical Skyrme model (see Eq. \eqref{constraint2}), however, these new relations are more complicated, since they involve high powers of the scale factor $h_0$. Since the exact form of these relations is not as relevant, we will not present them explicitly.

A relevant feature of these new solutions is that for exact solutions to exist, the sign of the coupling constants changes for different values of $N$. Specifically, from Eq. \eqref{newconst} we see that for $N=2$, $c_8$ and $c_6$ must have different signs, while for $N \ge 3$, $c_6$ and $c_8$ have the same sign. In this scenario, as expected, $c_8$ is a small number compared to $c_6$, and the difference becomes larger as $N$ grows.

On the other hand, from the expression of the energy density\footnote{Here where we have used Eq. \eqref{newconst}.}
\begin{equation}
    \varepsilon = T_{tt} = \frac{3}{2} K a_N \frac{(h_0^2 + \lambda)}{h_0^4}  +\frac{1}{4} c_6 a_N \frac{1}{h_0^6} \ , 
\end{equation}
we see that this expression is positive definite, at least for any $c_6 > 0 $ (which is compatible with the values of $c_6$ considered in the literature \cite{Floratos}-\cite{Adam:2020yfv}).  

\subsection{Skyrmion at finite volume}

A relevant (and unexpected) feature of the self-gravitating skyrmions is that, from them, it is possible to construct skyrmion states in flat space-time (for detailed construction, see Refs. \cite{Eloy2} and \cite{Pedro1}). To achieve this, it is enough to consider the matter field in Eq. \eqref{U5}, including a soliton profile (to be solved) and considering a metric that describes a space of finite volume. Explicitly, the matter field $U$ is given by
\begin{gather} \label{U6}
 U(x^\mu)= e^{F_1(x^\mu) \cdot T_3} e^{F_2(x^\mu) \cdot T_2} e^{F_3(x^\mu) \cdot T_3}  \ , \\
F_1(x^\mu) = x \ , \qquad F_2(x^\mu) =  H(y) \ , \qquad F_3(x^\mu) = z \ ,  \notag
\end{gather}
while the space-time metric reads
\begin{equation} \label{box}
    ds^2 = -dt^2 + l^2 (dx^2+dy^2+dz^2) \ .
\end{equation}
Here we have renamed the coordinates to emphasize that we are now in flat space-time, but the range of the spatial coordinates remains the same. In Eqs. \eqref{U6} and \eqref{box}, $H$ is the profile of the skyrmion and $l$ is the size of the box in which the solitons are confined.

One can check that, by introducing the Ansatz in Eqs. \eqref{U6} and \eqref{box} into the generalized Skyrme equations in Eq. \eqref{Eqgeneralized}, the complete set of field equations is reduced to a single differential equation for the soliton profile. Even more, this is a first-order equation given by
\begin{align} \notag
   & (H')^2 + \frac{\lambda}{4(2l^2+\lambda)}\cos(2H) + \frac{c_6}{4K l^2(2l^2+\lambda)} \sin^2(H) (H')^2 \\ & +  \frac{c_8}{512 K l^4(2l^2+\lambda)}\{ \cos(4H) - 4 \cos(2H) +32 \sin^2(H)(H')^2 -96 \cos^2(H)(H')^4 \}  \,  = \,  E_0 \ . \label{Eqfinal}
\end{align}
The same value of the topological charge shown in Eq. \eqref{charge} is obtained when the following boundary conditions are imposed for the soliton profile: $H(2 \pi) = \pi/2$, $H(0)=0$. Then, the integration constant $E_0$ in Eq. \eqref{Eqfinal} will be fixed.

As Eq. \eqref{Eqfinal} is quartic and reducible to a quadrature; $(dH/dy)^2= \eta(H, E_0)$, the condition $\eta(H, E_0) \ge 0 $ leads to a bound for the integration constant, namely
\begin{equation*}
    E_0 \ge -  \frac{\sec^2(H)}{3072 c_8 K l^4(2l^2+\lambda)} \{ \mathrm{C}_1 + \mathrm{C}_2 \cos(2H) + \mathrm{C}_3 \cos(4H) + \mathrm{C}_4 \cos(6H) \} \ , 
\end{equation*}
where the coefficients $\mathrm{C}_i$ are 
\begin{align*}
   \mathrm{C}_1 \, = \, & -8 l^2 \left(6 c_6^2 l^2+c_6 \left(3 c_8+64 K l^4 \left(\lambda +2 l^2\right)\right)+4 K l^2 \left(c_8 \left(7 \lambda +8 l^2\right)+64 K l^4
   \left(\lambda +2 l^2\right)^2\right)\right) \ , \\ 
   \mathrm{C}_2 \, = \, & 32 c_8 l^2 \left(c_6+8 K l^4-2 \lambda  K l^2\right)+64 c_6 l^4 \left(c_6+8 K l^2 \left(\lambda +2 l^2\right)\right)+\frac{37 c_8^2}{4}  \ , \\ 
   \mathrm{C}_3 \, = \, & -16 c_6^2 l^4-8 c_8 l^2 \left(c_6+12 \lambda  K l^2\right)+\frac{c_8^2}{2} \ ,  \\ 
   \mathrm{C}_4 \, = \, & -\frac{3 c_8^2}{4} \ . 
\end{align*}
The present construction is the natural generalization of the results reported in Ref. \cite{Pedro1}, where the first exact solution of the Skyrme model in flat space-time was found. Here we have included the higher-order correction in the 't Hooft expansion and arbitrary flavors, allowing high topological charge without any extra parameter. It is important to mention that this solution can be generalized to consider different lengths of the cavity in which the solitons are confined, as well as adding some free parameters in the matter field (which are allowed in the flat space-time case). Also, it is possible to promote the soliton profile to a time-dependent one. The generalization of this solution, together with their physical properties and consequences, will be explored in a forthcoming paper.

\section{Conclusions} \label{sec-4}

We have shown that the generalized Einstein $SU(N)$-Skyrme model admits exact solutions describing self-gravitating multi-skyrmions. These configurations are the natural generalization of the configurations reported in Refs. \cite{Eloy1} and \cite{Eloy2} for the cases with baryon number $B=1$ and $B=4$, respectively. Noticing that the Skyrme field can be easily written in the Euler angles parametrization, we have generalized the solutions in two different ways. First, using the maximal embedding Ansatz of $SU(2)$ into $SU(N)$, we have constructed solutions for any value of the flavor number, which allows the existence of skyrmion states with arbitrary high baryon number. Second, considering higher-order corrections to the Skyrme model that comes from the 't Hooft expansion, we have shown that exact solutions also exist when fine-tuning between the new coupling constants are fulfilled. With the above results, we can conclude that the self-gravitating skyrmions possess a kind of universal character, since the solutions are robust to the increase of the color and flavor numbers. These results open the possibility of modeling compact objects with high baryonic charge, such as neutron stars, in a completely analytical way. Finally, the resolution methods introduced in this paper can be applied to the construction of traversable wormholes and cosmological solutions, such as those explored in Refs. \cite{Eloy1}, \cite{Eloy2} and \cite{Pavluchenko:2016gyd}-\cite{Canfora:2019sxn}.

\acknowledgments

The author thanks Fabrizio Canfora and Marcela Lagos for many insightful comments and discussions.


\end{document}